\documentclass[aps,pre,showpacs,amsmath,amssymb,amsfonts,superscriptaddress,lengthcheck,twocolumn,longbibliography]{revtex4-1}

\usepackage{color}
\usepackage{graphicx}
\usepackage{amsthm}
\usepackage{subfigure}
\usepackage{tabu}
\usepackage{verbatim}
\usepackage{dcolumn}
\usepackage{bm}
\usepackage{epsf}
\usepackage{color}
\usepackage[colorlinks=true,citecolor=blue,linkcolor=blue,urlcolor=blue]{hyperref}
\usepackage{hhline}
\usepackage{mathtools}
\usepackage{bbold}

\newcommand{\be}{\begin{equation}}
\newcommand{\ee}{\end{equation}}
\newcommand{\ba}{\begin{align}}
\newcommand{\ea}{\end{align}}
\newcommand{\bi}{\begin{itemize}}
\newcommand{\ei}{\end{itemize}}

\newcommand{\la}{\left\langle}
\newcommand{\ra}{\right\rangle}
\newcommand{\pd}{\partial}

\newcommand{\bla}{bla\\bla\\bla\\bla\\bla}

\newcommand{\mb}[1]{\mbox{\boldmath$#1$}}
\newcommand{\mc}[1]{\mathcal{#1}}

\begin{document}

\title{Series expansion of the excess work using nonlinear response theory}

\author{Pierre Naz\'e}
\email{p.naze@ifi.unicamp.br}

\affiliation{\it Instituto de F\'isica `Gleb Wataghin', Universidade Estadual de Campinas, 13083-859, Campinas, S\~ao Paulo, Brazil}

\author{Marcus V. S. Bonan\c{c}a}
\email{mbonanca@ifi.unicamp.br}

\affiliation{\it Instituto de F\'isica `Gleb Wataghin', Universidade Estadual de Campinas, 13083-859, Campinas, S\~ao Paulo, Brazil}

\date{\today}

\begin{abstract}

\begin{center}
{\bf Abstract}
\end{center}

The calculation of observable averages in non-equilibrium regimes is one of the most important problems in statistical physics. Using the Hamiltonian approach of nonlinear response theory, we obtain a series expansion of the average excess work and illustrate it with specific examples of thermally isolated systems. We report the emergence of non-vanishing contributions for large switching times when the system is subjected to strong driving. The problem is solved by using an adapted multiple-scale method to supress these secular terms. Our paradigmatic examples show how the method is implemented generating a truncated series that obeys the Second Law of Thermodynamics.
\end{abstract}

\maketitle

\section{Introduction}
\label{sec:intro} 

Nonlinear response theory is one of the few alternatives to go beyond the linear response regime when exact solutions are absent. The derivation of the far-from-equilibrium behavior within the nonlinear response framework has been done considering both Hamiltonian \cite{kubo1957, tani1964, morita1967, peterson1967, samokhin1968, kenkre1971, samokhin1972, kenkre1973, murayama1981,evans1990} and stochastic \cite{hanggi1978,bouchaud2005,marconi2008,lippiello2008,villamaina2009,colangeli2011,mallick2011,basu2015,holsten2021} microscopic dynamics. Pure macroscopic considerations has been also pursued \cite{callen1959,glansdorff1971,maes2003,andrieux2007,stratonovich2012}. In this work, focusing in the Hamiltonian approach of the problem, we consider the non-equilibrium average of the generalized force that appears in the expression of the thermodynamic work
\be
W = \int_0^\tau \overline{\frac{\pd \mathcal{H}}{\pd \lambda}}\frac{\pd \lambda}{\pd t}dt,
\label{eq:work1}
\ee 
where $\mc{H}$ is the Hamiltonian of interest, $\lambda$ is an externally controlled parameter that varies in time and $\overline{(\cdot)}$ denotes the non-equilibrium average. By subtracting the corresponding quasistatic work from the previous expression, one defines the so-called excess work \cite{nieuwenhuizen2005,nieuwenhuizen2007,crooks2012,bonanca2015} whose value is expected to vanish as the time $\tau$, required to perform the change in $\lambda$, increases. Such constraint is imposed by the Second Law of Thermodynamics in either isothermal or adiabatic processes \cite{jarzynski2020}.

Equation \eqref{eq:work1} tells us that by expanding the non-equilibrium average of $\partial \mathcal{H}/\partial\lambda$, we will automatically have a series expansion of the thermodynamic work whose behavior has to agree with the above-mentioned constraint. However, it is well-known that perturbative approaches to time-dependent problems often lead to asymptotic series and present secular terms \cite{arnold2007,lichtenberg2013,logan2013}. To exemplify this ubiquitous behavior, consider the Duffing equation
\be
\ddot{u}+u = -\epsilon u^3,
\ee
which can be understood as a rescaled harmonic oscillator subjected to an external nonlinear force term. When $\epsilon\ll 1$, the solution could be expanded in a series on the parameter $\epsilon$
\be
u(t) = u_0(t)+\epsilon u_1(t) + \epsilon^2 u_2(t) + ... .
\label{eq:expansionamplitude}
\ee
Considering the initial conditions $u(0)=1$ and $\dot{u}(0)=0$, the solution expanded up to first-order becomes
\be
u(t) = \cos{t}+\epsilon\left[\frac{1}{32}(\cos{3t}-\cos{t})-\frac{3}{8}t\sin{t}\right],
\ee
which clearly diverges for large $t$, even if $\epsilon$ is small. 

It is well known that what produces such divergence is the fact that the perturbation affects not only the amplitudes but also the frequencies of the oscillatory functions appearing in the expansion \cite{arnold2007,lichtenberg2013,logan2013}. In some situations, the secular terms might appear in higher orders and the truncation of the series to its lowest orders may lead to a good approximation.  However, there are techniques in which the use of multiple-time scale strategies to eliminate the secular terms leads to a well-behaved series expansion \cite{arnold2007,lichtenberg2013,logan2013}. Along these lines, one of the first methods ever used in this kind of problem is the Lindstedt-Poincar\'e method \cite{lindstedt1882,poincare1893} in the context of classical perturbation theory in celestial mechanics \cite{lichtenberg2013,fasano2006}. The idea behind it is simple and can be rephrased as follows: besides the series expansion of the amplitude, as given by Eq. \eqref{eq:expansionamplitude}, the instant of time $t$ must be also rescaled according to
\be
\tau = t(\omega_0+\epsilon\omega_1+\epsilon^2\omega_2+...),
\ee  
where $\omega_{0}$ is the frequency of oscillation of the unperturbed system and the $\omega_{i}$, $i=1,2,...$, are not just frequency corrections but also help us to eliminate the secular terms. In this work, we will use an adapted Lindstedt-Poincar\'e method to systematically fix the series expansion of the excess work upto an arbitrary order. In other words, the method presented here supresses the emergence of secular terms and hence produces a meaningful series expansion that agrees with the Second Law of Thermodynamics.

This work is organized in the following way. In Sec. \ref{sec:nlrt}, we present the Hamiltonian approach of nonlinear response theory. In Sec. \ref{sec:eas}, we exemplify our method using the thermally isolated harmonic oscillator subjected to a strong linear driving in the stiffness parameter. We verify the emergence of secular terms in the series expansion obtained from nonlinear response theory and show how to suppress them employing the ideas of Lindstedt-Poincaré method, both in classical and quantum cases. In Sec. \ref{sec:final} we make our final considerations.

%%%%%%%%%%%%%%%%%%%%%%%%%%%%%%%%%%%%%%%%%%%%%%%%%%%%%%%%%%%%%%%%%%%%%

\section{Nonlinear response theory}
\label{sec:nlrt}

%%%%%%%%%%%%%%%%%%%%%%%%%%%%%%%%%%%%%%%%%%%%%%%%%%%%
%%%%%%%%%%%%%%%%%%%%%%%%%%%%%%%%%%%%%%%%%%%%%%%%%%%%

\subsection{Preliminaries}

Consider a classical system with a Hamiltonian $\mc{H}(\mb{z}(\mb{z_0},t),\lambda(t))$, where $\mb{z}$ is a point in the phase space $\Gamma$ evolved from the initial point $\mb{z}_0$ according to Hamiltonian dynamics of $\mc{H}$ and $\lambda(t)$ is a control parameter. Initially, this system is in contact with a heat bath of inverse temperature $\beta\equiv {(k_B T)}^{-1}$, where $k_B$ is Boltzmann's constant and its Hamiltonian is $\mc{H}_0 = \mc{H}(\mb{z_0},\lambda_0)$.After that, the system is decoupled from the initial heat bath and, during a switching time $\tau$, the control parameter is changed from $\lambda_0$ to $\lambda_0+\delta\lambda$. Given some observable $\vartheta(\mb{z_0},\lambda(t))$, our interest is to calculate its non-equilibrium average
\be
\overline{\vartheta}(t) = \int_{\Gamma}\vartheta(\mb{z}(\mb{z_0},t),\lambda(t))\rho(\mb{z}(\mb{z_0},t))d{\mb{z_0}},
\label{eq:totalresp}
\ee
where $\rho(\mb{z}(\mb{z_0},t))$ is the non-equilibrium phase-space distribution of the system. Such quantity evolves according to the Liouville equation
\be
\dot{\rho} = -\{\rho,\mc{H}\} \equiv {\mc L}\rho, \quad \rho(\mb{z_0},0) = \rho_c(\mb{z_0})
\label{eq:liouville}
\ee
where $\{\cdot,\cdot\}$ is the Poisson bracket, ${\mc L}$ the Liouville operator, and $\rho_c:=\exp{(-\beta\mathcal{H}_{0})}/\mc{Z}$ is the canonical ensemble, where $\mc{Z} = \int_{\Gamma}\exp{(-\beta\mathcal{H}_{0})}d\mb{z_0}$. We remark that the Liouville operator is regarding only the system of interest. In the particular case where the Hamiltonian $\mc{H}=\mc{H}_0$, the formal solution of the Liouville equation is
\be
\rho(\mb{z}(\mb{z_0},t)) = e^{t\mathcal{L}_0}\rho_c(\mb{z_0}),
\ee
where ${\mc L}_0$ is the Liouville operator associated to $\mc{H}_0$. We remark also that the time evolution operator $e^{-t\mathcal{L}_0}$ can act on observables $\vartheta(\mb{z_0})$, indicating how they evolve in time
\be
\vartheta(\mb{z}(\mb{z_0},t)) = e^{-t\mathcal{L}_0}\vartheta(\mb{z_0}),
\ee
where $\mb{z}(\mb{z_0},t)$ is the solution of Hamilton equations considering the Hamiltonian $\mc{H}_0$. 

The control parameter can be expressed as
\be
\lambda(t) = \lambda_0+g(t)\delta\lambda,
\ee
where the protocol $g(t)$ must satisfy the following boundary conditions
\be
g(0)=0,\quad g(\tau)=1. 
\label{eq:bc}
\ee
We also consider that $g(t)\equiv g(t/\tau)$, which means that the time intervals are measured in units of the switching time $\tau$. Additionally, the quantity $\delta\lambda/\lambda_{0}$ quantifies the driving strength of the process. 

In linear response theory, the response of the observable is restricted to first-order in $\delta\lambda/\lambda_{0}$. This can be achieved by expanding the instantaneous observable and the non-equilibrium ensemble in a power series of $\delta\lambda/\lambda_{0}$, regroupping the corresponding terms and keeping only the first-order term. In nonlinear response theory, the procedure is the same, but the truncation goes further than that used in linear response theory. Our goal is to calculate the average work performed by the switch of the control parameter $\lambda$ beyond the linear-response regime. Its expression is given by Eq. \eqref{eq:work1} and requires a series expansion of the observable $\partial \mathcal{H}/\partial\lambda$. In the following, we develop a systematic way of collecting the different orders given by non-linear response theory. This method will be exemplified in Sec. \ref{sec:eas}.

%%%%%%%%%%%%%%%%%%%%%%%%%%%%%%%%%%%%%%%%%%%%%%%%%%%%%%%%%
%%%%%%%%%%%%%%%%%%%%%%%%%%%%%%%%%%%%%%%%%%%%%%%%%%%%%%%%%

\subsection{Expansion in higher-order terms}
\label{subsec:se}

Our first step is to rewrite Eq. (\ref{eq:totalresp}) as a power series of the driving strength 
\be
\overline{\vartheta}(t) = \sum_{n=0}^{\infty}{\overline{\vartheta}}_{n}(t)\left(\frac{\delta\lambda}{\lambda_0}\right)^n.
\label{eq:obsexp}
\ee
To find the functions $\overline{\vartheta}_{n}(t)$, consider then the following expansions
\be
\vartheta(\mb{z}(\mb{z_0},t),\lambda(t)) = \sum_{n=0}^{\infty}\vartheta_n({\bf z_0},t)\left(\frac{\delta\lambda}{\lambda_0}\right)^n,
\label{eq:var_ensemble}
\ee 
and
\be
\rho(\mb{z}(\mb{z_0},t)) = \sum_{n=0}^{\infty}\rho_n({\bf z_0},t)\left(\frac{\delta\lambda}{\lambda_0}\right)^n.
\label{eq:exp_ensemble}
\ee
Knowing the observable $\vartheta(\mb{z}(\mb{z_0},t),\lambda(t))$, the functions $\vartheta_n({\bf z_0},t)$ are the coefficients of the Taylor expansion of $\vartheta$ in the control parameter $\lambda$ around $\lambda_0$
\be
\vartheta_n({\bf z_0},t) = \lambda_0^n \frac{g^n(t)}{n!}\frac{\partial^n \vartheta}{\partial\lambda^n}\Bigr|_{\lambda=\lambda_0},
\label{eq:varthetan}
\ee
which have been multiplied by $\lambda_0^n$ to keep the correct dimensions in Eq. \eqref{eq:var_ensemble}. We remark that when the observable depends on the control parameter, each term of the expansion is an instantaneous response due to the variation of such quantity. 

The terms $\rho_n({\bf z_0},t)$ are much more involved to obtain. They require a manipulation of Eq. \eqref{eq:liouville}, which is explained in the following. First, we consider the expansion of the Hamiltonian as a power series on the driving strength
\be
\mathcal{H} = \sum_{n=0}^{\infty}\mathcal{H}_n(\mb{z}(\mb{z_0},t))g^n(t)\left(\frac{\delta\lambda}{\lambda_0}\right)^n,
\label{eq:exp_hamiltonian}
\ee
whose coefficients are given by
\be
\mathcal{H}_n(\mb{z}(\mb{z_0},t)) = \frac{\lambda_0^n}{n!}\frac{\partial^n\mathcal{H}}{\partial\lambda^n}\Bigr|_{\lambda=\lambda_0}.
\ee 
We can construct a Liouville operator $\mathcal{L}_n$ for each $\mathcal{H}_n$
\be
\begin{aligned}
\mathcal{L}(\bullet) &= -\{\bullet,\mathcal{H}\} \\
&= \sum_{n=0}^{\infty}g^n(t)\left(\frac{\delta\lambda}{\lambda_0}\right)^n\left[-\{\bullet,\mathcal{H}_n\}\right] \\
&\equiv \sum_{n=0}^{\infty}g^n(t)\left(\frac{\delta\lambda}{\lambda_0}\right)^n\mathcal{L}_n(\bullet).
\end{aligned}
\ee 
In particular, we distinghish those Liouville operators which depend on time from the time-independent term,
\be
\mc{L} = \mc{L}_0+ \sum_{n=1}^{\infty}\mathcal{L}_n g^n(t)\left(\frac{\delta\lambda}{\lambda_0}\right)^n \equiv \mc{L}_0+\mc{L}_{\text{ext}}(t).
\ee
For the case where the initial ensemble is the canonical one, the integral form of the Liouville equation reads \cite{kubo1985}
\begin{multline}
\rho(\mb{z}(\mb{z_0},t),t) = e^{t\mathcal{L}_0}\rho_c({\bf z_0}) \\
+ \int_0^t e^{(t-s)\mathcal{L}_0}\mc{L}_{\text{ext}}(s)\rho(\mb{z}(\mb{z_0},s),s)ds,
\end{multline}
which, using the Liouville theorem, becomes
\begin{multline}
\rho(\mb{z}(\mb{z_0},t),t) = \rho_c({\bf z_0}) \\
+ \int_0^t e^{(t-s)\mathcal{L}_0}\mc{L}_{\text{ext}}(s)\rho(\mb{z}(\mb{z_0},s),s)ds.
\label{eq:rhoperturbative}
\end{multline}
Expanding now $\rho$ and $\mathcal{L}$ in a power series on the driving strength, we have
\begin{multline}
\sum_{n=0}^{\infty}\rho_n({\bf z_0},t)\left(\frac{\delta\lambda}{\lambda_0}\right)^n = \rho_c({\bf z_0}) \\
+ \sum_{i=1}^\infty\sum_{j=0}^\infty\int_0^t e^{(t-s)\mathcal{L}_0}\mathcal{L}_i \rho_j({\bf z_0},s)g^{i}(s)ds\left(\frac{\delta\lambda}{\lambda_0}\right)^{i+j}.
\label{eq:rhoexp}
\end{multline}
Finally, regroupping the terms with the same powers $\left(\delta\lambda/\lambda_0\right)^n$, we finally have
\be
\begin{aligned}
\rho_0({\bf z_0},t) & = \rho_c({\bf z_0}),\\
\rho_n({\bf z_0},t) & = \sum_{k=0}^{n-1}\int_0^t e^{(t-s)\mathcal{L}_0}\mathcal{L}_{n-k}\rho_k({\bf z_0},s)g^{n-k}(s)ds,
\label{eq:rhon}
\end{aligned}
\ee 
for $n\ge 1$. Therefore, to find $\rho_n$ one needs to know the previous solutions $\rho_{n-k}$, for $ 1\le k \le n$. We remark that the $\rho_{n}$ cannot be considered individually as valid probability distributions since their integral over the phase space is zero \cite{kubo1985}.

Finally, using Eqs. (\ref{eq:varthetan}) and (\ref{eq:rhon}), the term ${\overline{\vartheta}}_{n}(t)$ can be rearranged as 
\be
{\overline{\vartheta}}_{n}(t) = \sum_{k=0}^{n}\int_{\Gamma} \vartheta_k({\bf z_0},t)\rho_{n-k}({\bf z_0},t)d{\bf z_0}.
\label{eq:solobsexp}
\ee
Therefore, the averaged observable becomes 
\be
{\overline{\vartheta}}(t) = \sum_{n=0}^{\infty}\left[\sum_{k=0}^{n}\int_{\Gamma} \vartheta_k({\bf z_0},t)\rho_{n-k}({\bf z_0},t)d{\bf z_0}\right] \left(\frac{\delta\lambda}{\lambda_0}\right)^n.
\label{eq:average}
\ee
In the particular case where the observable $\vartheta$ is independent of the control parameter $\lambda(t)$, its nonequilibirum average becomes
\be
{\overline{\vartheta}}(t) = \sum_{n=0}^{\infty}\left[\int_{\Gamma} \vartheta({\bf z_0})\rho_n({\bf z_0},t)d{\bf z_0}\right] \left(\frac{\delta\lambda}{\lambda_0}\right)^n.
\label{eq:average2}
\ee

%%%%%%%%%%%%%%%%%%%%%%%%%%%%%%%%%%%%%%%%%%%%%%%%%%%%%%%%
%%%%%%%%%%%%%%%%%%%%%%%%%%%%%%%%%%%%%%%%%%%%%%%%%%%%%%%%

\subsection{Calculating non-equilibrium averages of arbitrary order}
\label{subsec:seqsteps}

To calculate non-equilibrium averages of arbitrary order, one may proceed as follows:

\begin{enumerate}

\item Choose an order $n$ to truncate  Eq. (\ref{eq:obsexp});

\item For $0\le k\le n$, calculate $\vartheta_k$ using Eq. (\ref{eq:varthetan});

\item For $0\le k\le n$, calculate $\mc{H}_k$ using Eq. (\ref{eq:exp_hamiltonian});

\item Calculate the solution $\mb{z}(\mb{z_0},t)$ of $\mc{H}_0$;

\item For $0\le k\le n$, calculate $\rho_k$ using Eq. (\ref{eq:rhoexp});

\item For $0\le k\le n$, calculate ${\overline{\vartheta}}_k$ using Eq. (\ref{eq:solobsexp});

\item Calculate ${\overline{\vartheta}}$ using Eq. (\ref{eq:average}).

\end{enumerate}

In what follows, we will illustrate this procedure with specific examples.

%%%%%%%%%%%%%%%%%%%%%%%%%%%%%%%%%%%%%%%%%%%%%%%%%%%%%%%%%%%%%
%%%%%%%%%%%%%%%%%%%%%%%%%%%%%%%%%%%%%%%%%%%%%%%%%%%%%%%%%%%%%
%%%%%%%%%%%%%%%%%%%%%%%%%%%%%%%%%%%%%%%%%%%%%%%%%%%%%%%%%%%%%

\subsubsection{Recovering linear response theory}

Let us use the above-mentioned sequence of steps to recover the standard result of linear response theory, which corresponds to the truncation at $n=1$. We restrict ourselves to the case in which the observable of interest does not depend on the control parameter $\lambda(t)$
\be
\vartheta({\bf z_0},\lambda(t)) = \vartheta_0({\bf z_0}).
\ee
For the sake of simplicity, we also consider that the expansion of the Hamiltonian $\mathcal{H}$ ends up exactly at first order,
\be
\mc{H}({\bf z_0},\lambda(t)) = \mc{H}_0({\bf z_0})+\mc{H}_1({\bf z_0})g(t)\left(\frac{\delta\lambda}{\lambda_0}\right).
\label{eq:hamiltonfirst}
\ee
According to the procedure described previously, we only need the first two terms of Eq. \eqref{eq:exp_ensemble}, whose coefficients read
\be
\rho_0({\bf z_0},t) = \rho_c({\bf z_0}),
\ee
\be
\rho_1({\bf z_0},t) = \int_0^t e^{(t-s)\mathcal{L}_0}\mathcal{L}_1\rho_c({\bf z_0})g(s)ds.
\ee
The term ${\overline{\vartheta}}_0$ is given by
\be
{\overline{\vartheta}}_0(t) = \int_{\Gamma} \vartheta_0({\bf z_0})\rho_c({\bf z_0})d{\bf z_0} \equiv \la\vartheta_0\ra,
\ee
while the term ${\overline{\vartheta}}_1$ reads
\be
{\overline{\vartheta}}_1(t) = \int_{\Gamma} \vartheta_0({\bf z_0})\rho_1({\bf z_0},t)d{\bf z_0},
\ee
which can be rewritten as
\be
\begin{aligned}
\int_{\Gamma}& \vartheta_0({\bf z_0})\rho_1({\bf z_0},t)d{\bf z_0} = \\
&= \int_{\Gamma}\vartheta_0({\bf z_0})\left[\int_0^t e^{(t-s)\mathcal{L}_0}\mathcal{L}_1\rho_c({\bf z_0})g(s)ds\right]d{\bf z_0}\\
&= -\int_0^t\left[\int_{\Gamma}\mathcal{L}_1 e^{-(t-s)\mathcal{L}_0}\vartheta_0({\bf z_0}) \rho_c({\bf z_0})d{\bf z_0}\right]g(s)ds\\
&= \int_0^t\phi_1(t-s)g(s)ds,
\end{aligned}
\ee
where we have used the antisymmetric property of the Liouville operator \cite{kubo1985}. The term $\phi_1(t)$ is the so-called (first-order) response function, defined by
\be
\phi_1(t) \equiv \la \{e^{-t\mathcal{L}_0}\vartheta_0({\bf z_0}),\mc{H}_1({\bf z_0})\} \ra,
\ee
where the symbol $\la(...)\ra$ denotes the equilibrium average on $\rho_c$. Therefore the non-equilibrium average of the observable upto its first-order is
\be
{\overline{\vartheta}}(t) = \la\vartheta_0\ra+\frac{\delta\lambda}{\lambda_0}\int_0^t\phi_1(t-s)g(s)ds,
\label{eq:standardlrt}
\ee
which is the standard result of linear response theory.

\subsubsection{Recovering the second-order}

To illustrate the calculation of the second-order term, we restrict ourselves again to the situation in which the observable does not depend on the control parameter and the Hamiltonian is exactly given by Eq. \eqref{eq:hamiltonfirst} . In this case, the non-equilibrium average of $\vartheta$ upto second order is given by Eq. \eqref{eq:standardlrt} plus the following term multiplied by $\left(\delta\lambda/\lambda_0\right)^2$
\be
{\overline{\vartheta}}_2(t) = \int_{\Gamma} \vartheta({\bf z_0})\rho_2({\bf z_0},t)d{\bf z_0},
\ee
where, according to Eq. \eqref{eq:rhon}, we have
\be
\begin{aligned}
\rho_2({\bf z_0},t) =\int_0^t\int_0^{s_1} & e^{(t-s_1)\mathcal{L}_0}\mathcal{L}_{1} e^{(s_1-s_2)\mathcal{L}_0}\mathcal{L}_{1}\rho_0({\bf z_0})\\
&g(s_2)g(s_1)ds_2ds_1.
\end{aligned}
\ee
Using the antisymmetric property of the Liouville operators as we did in the linear response case, we arrive at
\be
{\overline{\vartheta}}_2(t) = \int_0^t\int_0^{s_1}\phi_2(t-s_1,s_1-s_2)g(s_1)g(s_2)ds_1 ds_2,
\label{eq:2ndorder}
\ee
where
\be
\phi_2(t,s) = \langle \{e^{-s\mathcal{L}_0}\{e^{-t\mathcal{L}_0}\vartheta_0({\bf z_0}),\mc{H}_1({\bf z_0})\},\mc{H}_1({\bf z_0})\} \rangle,
\ee
is the second-order response function \cite{kubo1957}.

\vspace{0.3cm}

We remark that, even in higher orders, the nonequilibrium average can always be expressed in terms of response functions that depend only on equilibrium correlation functions. In other words, this is not an exclusive feature of linear response theory, but a consequence of the perturbative method used in Eq.\eqref{eq:rhoperturbative} with a zero-order term given by an equilibrium ensemble. 

%%%%%%%%%%%%%%%%%%%%%%%%%%%%%%%%%%%%%%%%%%%%%%%%%%%%%%%%%%%%%%%

\section{Series expansion of the excess work}
\label{sec:eas}

In what follows, we compare the exact solution of a simple but relevant example and the corresponding perturbative expression from non-linear response theory. Both results are obtained for the excess thermodynamic work performed along a finite-time process whose duration or switching time we denote by $\tau$ \cite{bonanca2015}. This comparison will show that the perturbative expansion described in the Sec.\ref{sec:nlrt} leads to an asymptotic series. However, the divergences of higher-order terms can be removed by the application of a modified Lindstedt-Poincar\'e method \cite{lindstedt1882,poincare1893}.

%%%%%%%%%%%%%%%%%%%%%%%%%%%%%%%%%%%%%%%%%%%%%%%%%%%%%%%%%%
%%%%%%%%%%%%%%%%%%%%%%%%%%%%%%%%%%%%%%%%%%%%%%%%%%%%%%%%%%

\subsection{Thermally isolated harmonic oscillator}
\label{subsec:tiho}

\begin{figure}
\centering
    \includegraphics[scale=0.55]{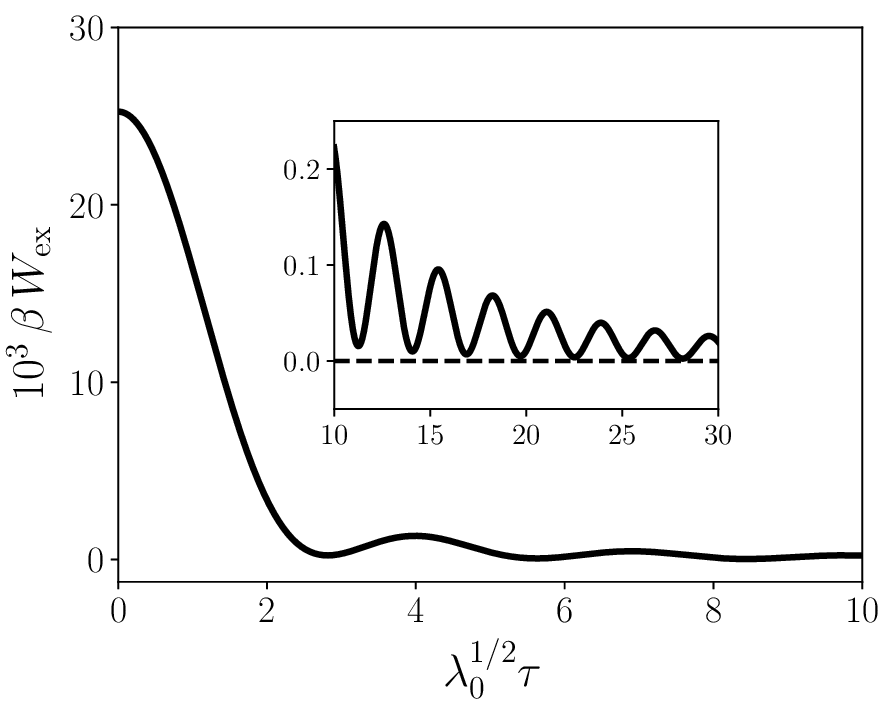}
    \caption{(Color online) Excess work for different switching times for the exact time-dependent dynamics given by the Hamiltonian (\ref{eq:tdhamilton}) and the protocol $\lambda(t)$ of Eq. (\ref{eq:protocol}). The curve was obtained from the analytical solution of Hamilton equations (see Appendix \ref{app:B}). The driving strength $\delta\lambda/\lambda_{0}$ was chosen equal to 0.5.}
\label{fig:1}
\end{figure}

We will apply the results of Sec. \ref{sec:nlrt} to a thermally isolated harmonic oscillator whose Hamiltonian is given by
\be
\mc{H}(\lambda(t)) = \frac{1}{2}\left[p^2+\lambda(t)q^2\right],
\label{eq:tdhamilton}
\ee
where the time-depedent stiffness $\lambda(t)$ is the control parameter, which is driven according to the linear protocol
\be
\lambda(t) = \lambda_0+\delta\lambda \frac{t}{\tau}.
\label{eq:protocol}
\ee
The oscillator is initially in equilibrium with a heat bath of inverse temperature $\beta = (k_B T)^{-1}$, which is removed before the starting of the non-equilibrium driving. Hence, the dynamics of the oscillator is Hamiltonian during the whole process. The average work performed on the system along the process was defined in Eq. \eqref{eq:work1}, where the term $\overline{\pd \mc{H}}/{\pd \lambda}$ is interpreted as a generalized force.  For thermally isolated systems, the energetic cost due to finite-time driving can be measured by the excess work,
\be
W_{\text{ex}} \equiv W-W_{\text{qs}},
\label{eq:wirr}
\ee
defined as the difference between the thermodynamic work, given by Eq. \eqref{eq:work1}, and the quantity $W_{\rm qs}$. The term $W_{\text{qs}}$ is the quasistatic work, whose value is the difference between the final and initial average energies of the system when it is driven along a quasistatic process \cite{jarzynski2020}. This quantity is obtained from the adiabatic invariant \cite{fasano2006} which, for the one-dimensional harmonic oscillator, is nothing but the action or, equivalently, the area in phase space enclosed by the curve of constant energy. In Appendix \ref{app:A}, we calculate the quasistatic work for our system, which is given by
\be
W_{\text{qs}} = \frac{1}{\beta}\left[\left(1+\frac{\delta\lambda}{\lambda_0}\right)^{1/2}-1\right].
\ee
Figure \ref{fig:1} depicts the excess work performed along the protocol \eqref{eq:protocol} for different switching times. Each point of the curve shown in Fig. \ref{fig:1} corresponds to the average of the difference between final and initial values of the Hamiltonian \eqref{eq:tdhamilton}. These averages are taken over the initial canonical ensemble. For the linear protocol (\ref{eq:protocol}), Hamilton equations are analytically solvable (see Appendix \ref{app:B}). We observe that the excess work vanishes for large $\tau$, in agreement with the Second Law of Thermodynamics. 

In what follows, we obtain a perturbative expression for $W_{\rm ex}$ based on non-linear response theory and verify that the decay for large $\tau$ is not observed for higher-order terms. We denote by $W_{\text{ex}}^{(n)}$ the $n$th contribution in the series expansion of the excess work.

%%%%%%%%%%%%%%%%%%%%%%%%%%%%%%%%%%%%%%%%%%%%%%%%%%%%%%%%%%%%%%%%%%%%%%%%%%%%
%%%%%%%%%%%%%%%%%%%%%%%%%%%%%%%%%%%%%%%%%%%%%%%%%%%%%%%%%%%%%%%%%%%%%%%%%%%%

\subsection{Perturbative expression via non-linear response theory}

To obtain a perturbative expression for $W_{\text{ex}}$ using non-linear response theory, we apply the sequence of steps described in Sec. \ref{subsec:seqsteps}. Following the definitions established in Sec. \ref{subsec:se} (see Eq. (\ref{eq:exp_hamiltonian})), the Hamiltonian (\ref{eq:tdhamilton}) under the protocol \eqref{eq:protocol} can be split into two terms
\be
\mathcal{H}_0 = \mathcal{H}|_{\lambda=\lambda_0}= \frac{1}{2}\left[p^2+\lambda_0 q^2\right],
\ee
\be
\mathcal{H}_1 = \lambda_0\frac{\pd\mathcal{H}}{\pd\lambda}\Big|_{\lambda=\lambda_0}= \lambda_0 \frac{q^2}{2}.  
\ee
The expansion of the excess work goes through the expansion of the generalized force, which is an average of the following observable
\be
\frac{\pd\mc{H}}{\pd \lambda} = \frac{q^2}{2}.
\ee 
The calculation of the $(n+1)$th term of the excess work is based on finding the response functions $\phi_n$
\begin{multline}
\phi_n(t_1,...,t_n) = \la\{e^{-t_1\mc{L}_0}\{...\{e^{-t_{n-1}\mc{L}_0}\times\right.\\
\left.\{e^{-t_n\mc{L}_0}\vartheta_0({\bf z_0}),\mc{H}_1({\bf z_0})\},\mc{H}_1({\bf z_0})\} ...,\mc{H}_1({\bf z_0})\}\ra,
\end{multline}
which are going to be used to calculate the $n$th term of the generalized force, as exemplified in Eq. \eqref{eq:2ndorder}.

To calculate those terms, one can implement the sequence of steps described in Sec. \ref{subsec:seqsteps} as a computer algorithm to obtain response functions of arbitrary order. In Appendix \ref{app:C} we deduce via mathematical induction the expansion of the quasistatic work, which is necessary as well to find the $(n+1)$th term of the excess work. 

\begin{figure}
\centering
    \includegraphics[scale=0.47]{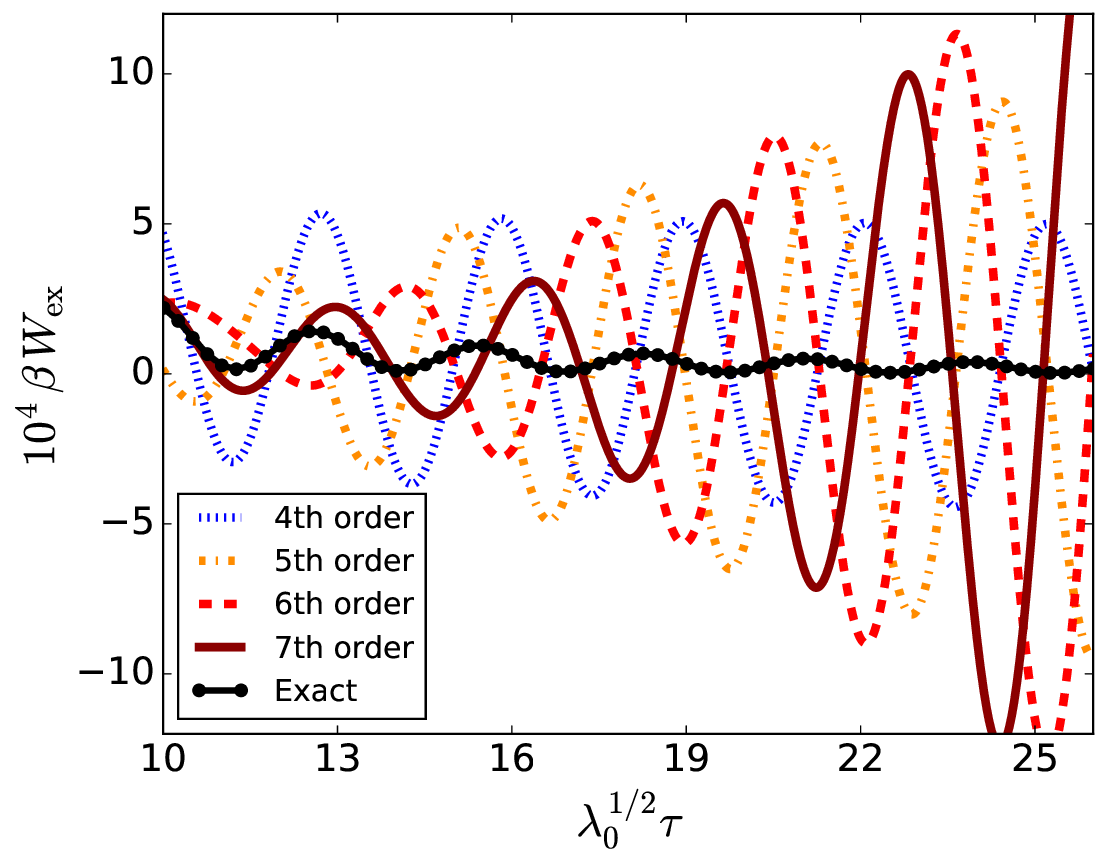}
    \caption{(Color online) Large $\tau$ behavior of the perturbative expansion of the excess work upto the $n$th order, with $4 \le n \le 7$. The results are compared with the exact result (see Fig. \ref{fig:1}). The driving strength was taken as $\delta\lambda_{0}/\lambda_{0}=0.5$.}
\label{fig:2}
\end{figure}

\begin{figure}
\centering
    \includegraphics[scale=0.47]{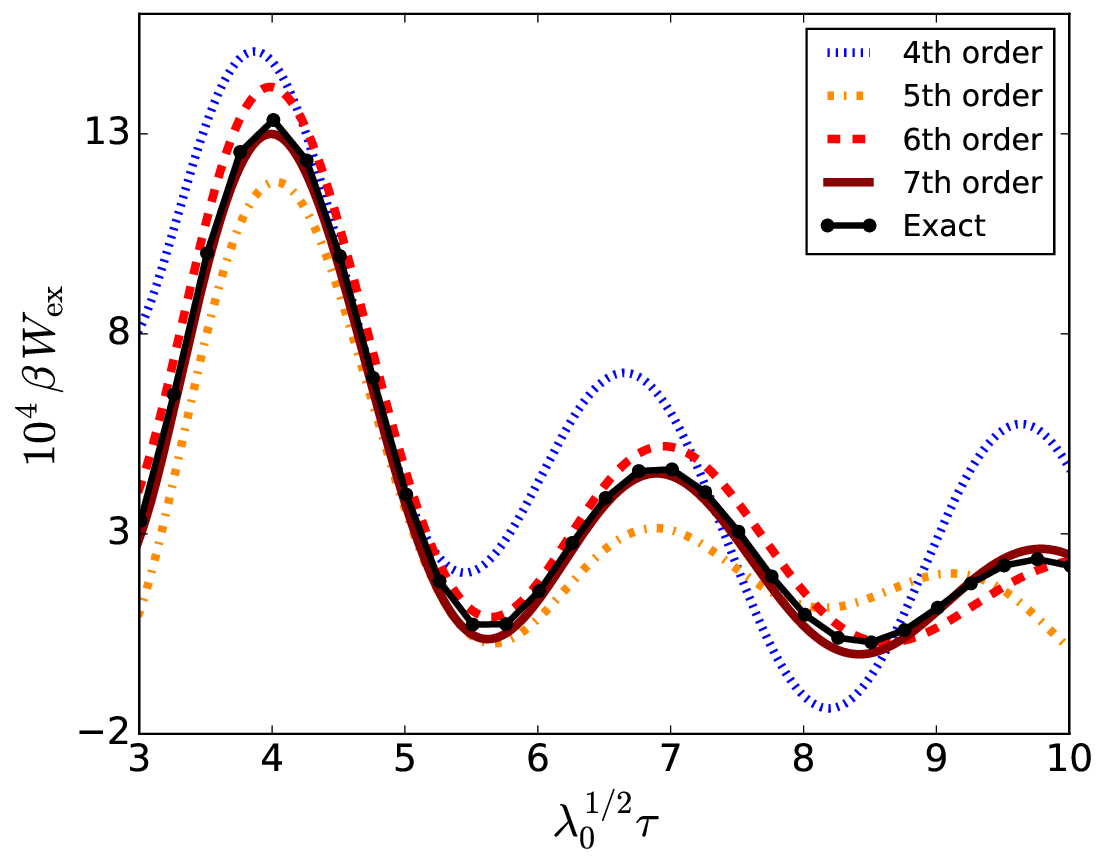}\\
    \caption{(Color online) Small $\tau$ behavior of the perturbative expansion of the excess work upto the $n$th order, with $4 \le n \le 7$. The results are compared with the exact result (see Fig. \ref{fig:1}). The driving strength was taken as $\delta\lambda_{0}/\lambda_{0}=0.5$.}
\label{fig:2.5}
\end{figure}

It is then straightforward to verify the emergence of secular terms for large $\tau$, something that is in stark contrast with the exact result (see Fig. \ref{fig:1}). For instance, the fourth-order term is already non-vanishing for large $\tau$,
\be
W_{\text{ex}}^{(4)} = ... + \frac{1}{\beta}\frac{\cos{(2\lambda_0^{1/2}\tau)}}{128}{\left(\frac{\delta\lambda}{\lambda_0}\right)}^4+...,
\label{eq:nonvanishing}
\ee
and the fifth-order term clearly diverges in the same limit
\be
W_{\text{ex}}^{(5)} = ... - \frac{1}{\beta}\frac{\lambda_0^{1/2}\tau\sin{(2\lambda_0^{1/2}}\tau)}{768}{\left(\frac{\delta\lambda}{\lambda_0}\right)}^5+... .
\ee
We depict in Fig. \ref{fig:2} the perturbative expansion of the excess work calculated upto the 7th order. We observe that the result starts to diverge after $n=4$ for large $\tau$. However, for relatively small values of $\tau$, the agreement with the exact result improves as higher-order terms are included in the sum (see Fig. \ref{fig:2.5}). 

The method we will present in the next section furnishes a new series in which the secular terms are suppressed. This reformulation of the expansion provided by nonlinear response theory will then be shown to successfully achieve a meaningful physical behavior for large values of $\tau$.   

%%%%%%%%%%%%%%%%%%%%%%%%%%%%%%%%%%%%%%%%%%%%%%%%%%%%%%%%%%%%%%%%%%%
%%%%%%%%%%%%%%%%%%%%%%%%%%%%%%%%%%%%%%%%%%%%%%%%%%%%%%%%%%%%%%%%%%%

\subsection{Supressing secular terms}
\label{subsec:sat}

The appearance of secular terms in a series expansion can be solved by using the Lindstedt-Poincar\'e technique. The main idea is to consider that the switching time can be written as a series in the driving strength
\be
\tau = \tau'\sum_{n=0}^\infty \frac{a_n}{n!} {\left(\frac{\delta\lambda}{\lambda_0}\right)}^n,
\label{eq:rescaletau}
\ee
where $\tau'$ is the rescaled switching time. The coefficients $a_n$ are obtained by demanding the removal of all divergent or non-vanishing terms (for large $\tau$) at each order. This becomes possible once Eq. (\ref{eq:rescaletau}) is inserted in the perturbative expansion obtained from nonlinear response theory and all functions of $\tau$ are expanded in powers of $\delta\lambda/\lambda_{0}$. Then, new coefficients for each order have to be found and the coefficients $a_{n}$ can be determined. Once the coefficients $a_n$ are determined (upto a certain order), the whole power series is rearranged. Below, we describe with some detail how the coefficients $a_0, a_1, a_2$, and $a_3$ are obtained.

The series expansion of $W_{\rm ex}$ starts with a second-order term that is well-behaved,
\be
W_{\text{ex}}^{(2)} = \frac{\sin^2\left(a_0 \lambda_0^{1/2} \tau' \right)}{8 a_0^2 \beta  \lambda_0 \tau'^2}\left( \frac{\delta \lambda}{\lambda _0}\right)^2.
\ee
As mentioned before, the first secular terms show up in $W^{(4)}_{\rm ex}$. To remove them, we plug Eq. \eqref{eq:rescaletau} truncated upto second order in the expressions of $W^{(2)}_{\rm ex}$, $W^{(3)}_{\rm ex}$ and $W^{(4)}_{\rm ex}$. We then expand all the terms upto the fourth-order and obtain the following combination of secular terms
\begin{multline}
W_{\text{ex}}^{(4)} = ...-\frac{a_1^2 \sin ^2\left(a_0 \lambda_0^{1/2} \tau' \right)}{8 a_0^2 \beta }+\frac{a_1^2 \cos^2\left(a_0\lambda_0^{1/2} \tau' \right)}{8 a_0^2 \beta }\\
+\frac{a_1 \cos\left(2 a_0\lambda_0^{1/2} \tau' \right)}{16 a_0 \beta }+\frac{\cos \left(2 a_0 \lambda_0^{1/2} \tau' \right)}{128\beta }+...,
\end{multline}
First, we observe that $a_0=1$, since $\tau$ is not rescaled at zeroth order. Thus these terms can be suppressed if we choose $a_{1}=-1/4$. This is our first example of how to obtain the coefficients of Eq. \eqref{eq:rescaletau}.

\begin{figure}
\centering
    \includegraphics[scale=0.47]{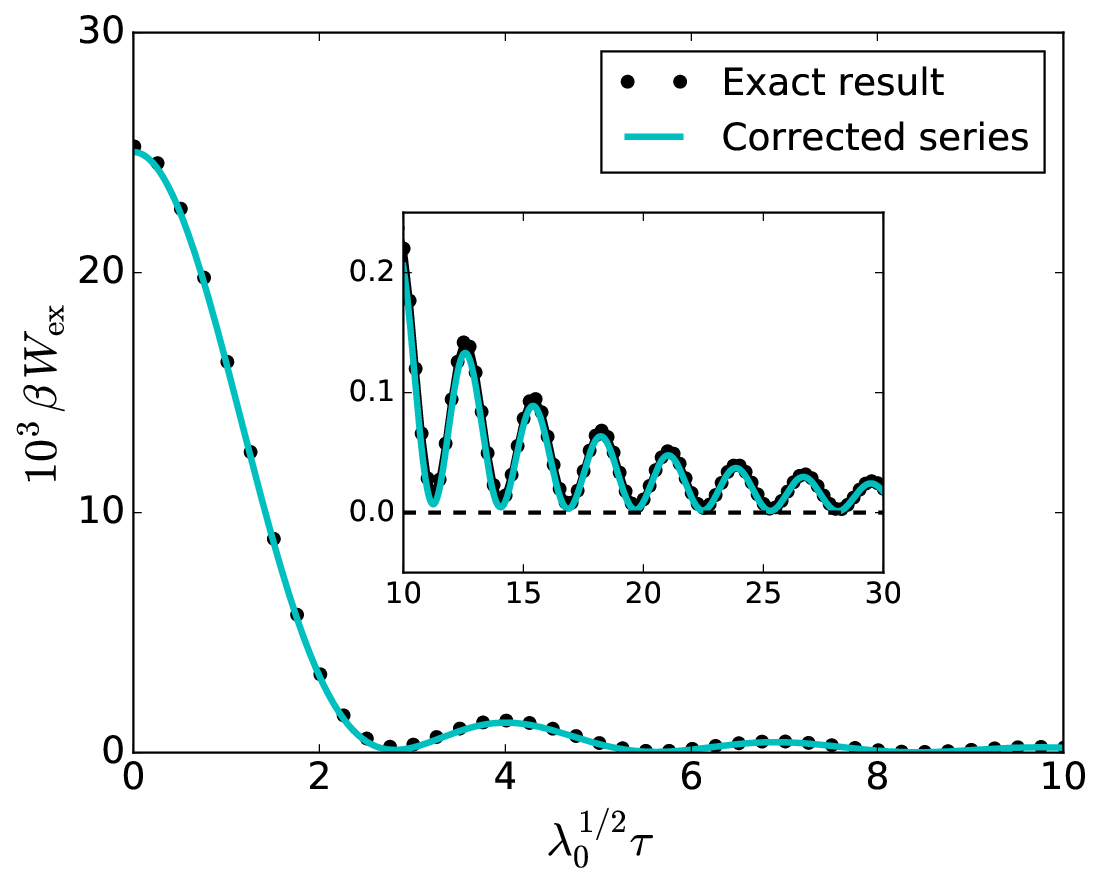}
    \caption{(Color online) Comparison between the exact result (black dots) and the corrected series expansion (blue solid line) of the excess work upto fifth order (see Eq. \ref{eq:wex5order}). It was used a driving strength of $\delta\lambda/\lambda_0=0.5$.}
\label{fig:3}
\end{figure}

The expansion mentioned above upto the fourth-order also involves the coefficient $a_{2}$. However, its value is determined only when we evaluate the series expansion upto the sixth order and demand that the secular terms appearing there vanish. Analogously, we have to go to the 9th order to determine the value of $a_{3}$. This procedure yields to
\be
a_0=1, \quad a_1=-\frac{1}{4},\quad a_2=\frac{5}{24},\quad a_3=-\frac{5}{16}.
\label{eq:coeffcients}
\ee
Substituting those values in the expression of the excess work expanded with Eq. \eqref{eq:rescaletau}, we produce a well-behaved series of the excess work upto 5th order
\begin{multline}
W_{\text{ex}}^{(5)}=\frac{\sin ^2(\lambda_0^{1/2}\tau' )}{8 \beta\lambda_0\tau'^2}{\left(\frac{\delta\lambda}{\lambda_0}\right)}^2-\frac{\sin ^2(\lambda_0^{1/2}\tau' )}{16 \beta\lambda_0\tau'^2}{\left(\frac{\delta\lambda}{\lambda_0}\right)}^3\\
+\left(\frac{77}{768 \beta\lambda_0\tau'^2}-\frac{13}{128 \beta\lambda_0^2\tau'^4}+\frac{17 \sin (2 \lambda_0^{1/2}\tau' )}{256\beta\lambda_0^{3/2}\tau'^3}\right.\\
\left.-\frac{23 \cos (2 \lambda_0^{1/2}\tau' )}{768 \beta\lambda_0\tau'^2}+\frac{13 \cos (2 \lambda_0^{1/2}\tau' )}{128\beta \lambda_0^2\tau'^4}\right){\left(\frac{\delta\lambda}{\lambda_0}\right)}^4\\
+\left(-\frac{69}{512\beta \lambda_0\tau'^2}+\frac{39}{256 \beta\lambda_0^2\tau'^4}-\frac{51 \sin (2\lambda_0^{1/2} \tau' )}{512 \beta\lambda_0^{3/2}\tau'^3}\right.\\
\left.+\frac{15 \cos (2 \lambda_0^{1/2}\tau' )}{512 \beta\lambda_0\tau'^2}-\frac{39 \cos(2\lambda_0^{1/2} \tau' )}{256\beta \lambda_0^2\tau'^4}\right){\left(\frac{\delta\lambda}{\lambda_0}\right)}^5.
\label{eq:wex5order}
\end{multline}
 Figure \ref{fig:3} presents a comparison between the corrected perturbative expression upto 5th order and the exact result. It was used a driving strength $\delta\lambda/\lambda_0=0.5$, which is about the maximum value before a deviation appears between both results. The value of $\tau'$ is calculated from Eq. \eqref{eq:rescaletau} using the coefficients $a_{i}$ obtained and the value of $\tau$ given as a boundary condition.

We remark that rescaling the switching time, whose value is one of the boundary conditions of the driving, is just a practical way of rescaling the frequency $\lambda_{0}^{1/2}$ since in all orders we have functions of the product $\lambda_{0}^{1/2}\tau$ \cite{lichtenberg2013,logan2013}.

\subsection{Extension to quantum case}

The method developed previously can easily be extended to the quantum case. Consider a quantum system with a Hamiltonian $\hat{\mathcal{H}}$. The first modification occurs at the Liouville equation, which becomes the Liouville-von Neumann equation
\be
\dot{\hat{\rho}} = \hat{\mathcal{L}}\hat{\rho}:=\frac{i}{\hbar}[\hat{\rho},\hat{H}],
\label{eq:vonneumann}
\ee 
where $\hat{\mathcal{L}}$ is the quantum Liouville operator, $\hat{\rho}$ is the density matrix and $[\cdot,\cdot]$ is the commutator. In the Heisenberg picture, the quantum Liouville operator evolves a operator $\hat{\vartheta}$ in time in the following way
\be
e^{-t\hat{\mathcal{L}}}\hat{\vartheta} := e^{i\frac{\hat{\mathcal{H}_0}t}{\hbar}}\hat{\vartheta}e^{-i\frac{\hat{\mathcal{H}_0}t}{\hbar}}.
\label{eq:quantumtime}
\ee 
Also the equilibrium average of an observable $\hat{\vartheta}$ is defined as
\be
\langle \hat{\vartheta}\rangle:={\rm Tr}(\hat{\vartheta}\hat{\rho_c}),
\ee
where $\hat{\rho_c}$ is the quantum canonical ensemble. 

The development of a power series in the quantum case is based on a procedure identical to the one developed in Sec. \ref{sec:nlrt}, but using the quantum Liouville operator. For example, for a quantum system of Hamiltonian
\be
\hat{\mathcal{H}} = \hat{\mathcal{H}}_0+\hat{\mathcal{H}}_1 g(t)\left(\frac{\delta\lambda}{\lambda_0}\right), 
\ee
the quantum analogues to the first and second response functions are
\be
\phi_1(t) = \frac{i}{\hbar}{\rm Tr}([e^{-t\hat{\mathcal{L}}}\hat{\vartheta}_0,\hat{\mathcal{H}_1}]),
\ee
\be
\phi_2(t) = -\frac{1}{\hbar^2}{\rm Tr}([e^{-t\hat{\mathcal{L}}}[e^{-t\hat{\mathcal{L}}}\hat{\vartheta}_0,\hat{\mathcal{H}_1}],\hat{\mathcal{H}_1}]).
\ee

We applied this extension to the quantum case to calculate the thermodynamic work given by Eq. \eqref{eq:work1} of the quantum analogue of Eq. \eqref{eq:tdhamilton}. The time-dependent Hamiltonian of such system is
\be
\hat{\mathcal{H}}(t) = \frac{\hat{p}^2}{2}+\left(\lambda_0+\frac{t}{\tau}\delta\lambda\right)\frac{\hat{q}^2}{2}, 
\ee
where $\hat{q}$ and $\hat{p}$ are the position and momentum operators respectively. The initial equilibrium situation was taken at temperature $T=0$, which means that all equilibrium averages were taken in the ground state of the unperturbed Hamiltonian. We used the package DiracQ from Mathematica to proceed in the calculations \cite{wright2013}.

\begin{figure}
\centering
    \includegraphics[scale=0.47]{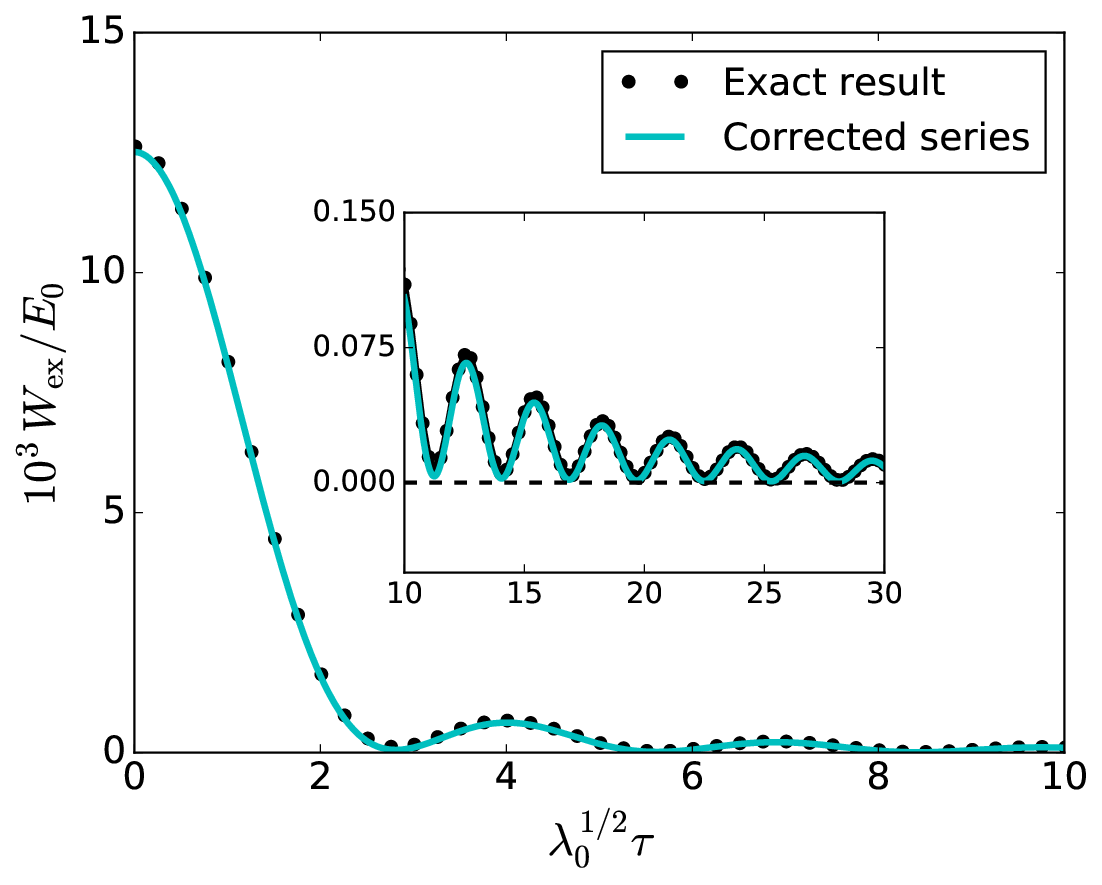}
    \caption{(Color online) Comparison between the exact result (black dots) and the corrected series expansion (blue solid line) of the excess work upto fifth order for a time-dependent quantum harmonic oscillator. It was used a driving strength of $\delta\lambda/\lambda_0=0.5$.}
\label{fig:4}
\end{figure}

In the present case, quantum nonlinear response theory leads to an asymptotic series of the excess work that is identical to the classical case with the only modification being
\be
\frac{1}{\beta} \rightarrow \frac{\hbar\sqrt{\lambda_0}}{2}:=E_0.
\ee
For instance, the leading order reads, 
\be
W_{\rm ex}^{(2)} =  \frac{\hbar\sin ^2(\lambda_0^{1/2}\tau' )}{16\lambda_0^{1/2}\tau'^2}{\left(\frac{\delta\lambda}{\lambda_0}\right)}^2.
\ee
Thus the asymptotic series are corrected by the same coeffcients given in Eq. \eqref{eq:coeffcients}. In Fig. \ref{fig:4} we compare the analytical result of the excess work with the corrected series given by the nonlinear response theory and the Lindstedt-Poincar\'e method. The fact that the quantum and classical series are the same (upto a different pre-factor) can be understood as a consequence of the existing analytical solution in both cases. In the case of a linear protocol for the stiffness parameter, it has been shown that Heisenberg's equations are identical to the classical case \cite{husimi1953,deffner2008}.

%%%%%%%%%%%%%%%%%%%%%%%%%%%%%%%%%%%%%%%%%%%%%%%%%%%%%%%%

\section{Final remarks}
\label{sec:final}

This work reports the existence of secular terms in the perturbative expansion of the excess work obtained from nonlinear response theory. To recover a meaningful result for large switching times, we employed a multiple-scale method that successfully suppresses these secular terms. This approach implies a rescaling of the frequency of oscillation observed in the behavior of the excess work. The implementation of such rescaling was made through an expansion of the switching time in powers of the driving strength. In classical canonical perturbation theroy, this is done only after the application of canonical transformations to action-angle variables which, in the context of nonequilibrium statistical physics, can be an unwanted step. Thus, our method has the advantage of keeping the formalism of nonlinear response theory unchanged. This rescaling of the frequency affects applications such as shortcuts to adiabaticity based on response theory \cite{acconcia2015} and possible future applications on optimization of the nonequilibrium work. We remark also that our approach can be used .to obtain the non-equilibrium behavior of other quantities such as the relative entropy \cite{vaikuntanathan2009}. Finally, we expect the Lindstedt-Poincar\'e method presented here to be useful also in the nonlinear response theory obtained from stochastic dynamics. Analogously to the Hamiltonian case, the starting point of the perturbative expansion is also the partial differential equation for the phase-space distribution \cite{hanggi1978,risken1996}.

\begin{acknowledgments}

P. N. gratefully thank Otavio L. Canton for the presentation which suggested the key to solve this problem. This work was financially supported for FAPESP (Funda\c{c}\~{a}o de Amparo \`a Pesquisa do Estado de S\~ao Paulo) (Brazil) (Grants No. 2018/06365-4, No. 2018/21285-7 and No. 2020/02170-4) and for CNPq (Conselho Nacional de Desenvolvimento Cient\'ifico e Pesquisa) (Brazil) (Grant No. 141018/2017-8).  

\end{acknowledgments}

\appendix

\section{Calculation of $W_{\rm qs}$}
\label{app:A}

The quasistatic work is given by the difference of the equilibrium averages of the energies calculated along the quasistatic process
\be
W_{\rm qs} = \la E_{\rm f}^{\rm ad}\ra -\la E_{\rm i}\ra,
\label{eq:wqsapp}
\ee
which are obtained by the conservation of the adiabatic invariant $\Omega(E, \lambda)$. This quantity is given by the area enclosed by the energy shell, which can be written as
\be
\Omega(E, \lambda) = \int_0^Ed\mathcal{H}\int_0^{2\pi} \mathcal{J}(\theta,\mathcal{H})d\theta,
\ee
where $\mathcal{J}$ is the Jacobian of the action-angle transformation $(q,p)\rightarrow (\theta,\mathcal{H})$. In the particular case of the harmonic oscillator, the adiabatic invariant is
\be
\Omega(E, \lambda) = \frac{2\pi E}{\sqrt{\lambda}}.
\ee
If the adiabatic invariant is conserved along the quasistatic process, it holds
\be
E_{\rm f}^{\rm ad}=E_{\rm i}{\left(\frac{\lambda_{\rm f}}{\lambda_{\rm i}}\right)}^{1/2}.
\label{eq:adbinv}
\ee
Using Eq. \eqref{eq:adbinv} on Eq. \eqref{eq:wqsapp}, the quasistatic work becomes
\be
W_{\text{qs}} = \frac{1}{\beta}\left[\left(\frac{\lambda_{\rm f}}{\lambda_{\rm i}}\right)^{1/2}-1\right].
\ee

\section{Solutions of Eq. (\ref{eq:tdhamilton})}
\label{app:B}

Given the initial conditions $q(0)=q_0$ and $p(0)=p_0$, the time-dependent Hamiltonian of Eq. (\ref{eq:tdhamilton}) has the following exact solution  

\begin{multline}
q(t) = \frac{\pi}{\sqrt[3]{\delta \lambda}} \left[\text{Ai}\left(-\frac{t \delta \lambda+\lambda_0 \tau }{\delta \lambda ^{2/3} \sqrt[3]{\tau}}\right) \left( \sqrt[3]{\tau }p_0\text{Bi}\left(-\frac{\lambda_0 \tau ^{2/3}}{\delta\lambda ^{2/3}}\right)\right.\right.\\
\left.+\sqrt[3]{\delta \lambda } q_0\text{Bi}'\left(-\frac{\lambda_0 \tau ^{2/3}}{\delta\lambda ^{2/3}}\right)\right)-\text{Bi}\left(-\frac{t \delta\lambda +\lambda_0 \tau }{\delta \lambda ^{2/3}\sqrt[3]{\tau }}\right)\times\\
\left.\left( \sqrt[3]{\tau }p_0\text{Ai}\left(-\frac{\lambda_0 \tau ^{2/3}}{\delta\lambda ^{2/3}}\right)+\sqrt[3]{\delta \lambda } q_0\text{Ai}'\left(-\frac{\lambda_0 \tau ^{2/3}}{\delta\lambda ^{2/3}}\right)\right)\right],
\end{multline}
\begin{multline}
p(t)=\frac{\pi}{\sqrt[3]{\tau }}\left[\text{Bi}'\left(-\frac{t \delta \lambda +\lambda_0 \tau }{\delta \lambda ^{2/3} \sqrt[3]{\tau}}\right) \left( \sqrt[3]{\tau }p_0 \text{Ai}\left(-\frac{\lambda_0 \tau ^{2/3}}{\delta \lambda^{2/3}}\right)\right.\right.\\
\left.+\sqrt[3]{\delta \lambda }q_0 \text{Ai}'\left(-\frac{\lambda_0 \tau ^{2/3}}{\delta\lambda ^{2/3}}\right)\right)-\pi  \text{Ai}'\left(-\frac{t \delta \lambda +\lambda_0 \tau }{\delta \lambda^{2/3} \sqrt[3]{\tau }}\right)\times\\ 
\left.\left( \sqrt[3]{\tau } p_0\text{Bi}\left(-\frac{\lambda_0 \tau^{2/3}}{\delta \lambda ^{2/3}}\right)+\sqrt[3]{\delta \lambda }q_0 \text{Bi}'\left(-\frac{\lambda_0\tau ^{2/3}}{\delta \lambda ^{2/3}}\right)\right)\right],
\end{multline}
where $\text{Ai}(x)$ and $\text{Bi}(x)$ are respectively the Airy functions of first and second type, which are defined as
\be
\text{Ai}(x) \equiv \frac{1}{\pi}\int_0^\infty \cos{\left(\frac{t^3}{3}+x t\right)} dt,
\ee
\be
\text{Bi}(x) \equiv \frac{1}{\pi}\int_0^\infty \left[\exp{\left(-\frac{t^3}{3}+x t\right)}+\sin{\left(\frac{t^3}{3}+x t\right)}\right] dt.
\ee

\section{Expansion of $W_{\text{qs}}$}
\label{app:C}

We are going to show that the quasistatic work 
\be
W_{\text{qs}} = \frac{1}{\beta}\left[\left(1+\frac{\delta\lambda}{\lambda_0}\right)^{1/2}-1\right]
\label{eq:wquasi}
\ee
can be expressed as
\be
W_{\text{qs}} = \frac{1}{\beta}\sum_{n=1}^\infty {(-1)}^{n+1}\frac{(2n-3)!!}{2^n n!}\left(\frac{\delta\lambda}{\lambda_0}\right)^n.
\label{eq:wqsexp}
\ee
Consider, by analogy with the function of Eq. \eqref{eq:wquasi}, the following function 
\be
f(x) = (1+x)^{1/2}-1.
\ee
Being $f(0)=0$, we state that the $n$-th derivative, for $n\ge 1$, is given by
\be
f^{(n)}(x) = (-1)^{n+1}\frac{(2n-3)!!}{2^n}{(1+x)}^{-\frac{2n-1}{2}}.
\label{eq:fder}
\ee
It is easy to see that the result holds for $n=1$. Taking the derivative of Eq. (\ref{eq:fder}), we have
\be
f^{(n+1)}(x)=(-1)^{(n+1)+1}\frac{(2(n+1)-3)!!}{2^{(n+1)}}{(1+x)}^{-\frac{2(n+1)-1}{2}},
\ee
which proves, by mathematical induction, our statement. As the Taylor expansion is given by
\be
f(x) = f(0)+\sum_{n=1}^\infty \frac{f^{(n)}(0)}{n!}x^n,
\label{eq:taylorexp}
\ee
Eq. (\ref{eq:wqsexp}) holds by applying Eq. (\ref{eq:fder}) in Eq. (\ref{eq:taylorexp}).

%\bibliographystyle{unsrt}
%\bibliography{ASNRref.bib}


\end{document}